\documentclass[twocolumn,showpacs,preprintnumbers,amssymb]{revtex4}
\usepackage{pdfpages}
\usepackage{graphics}
\usepackage{epsfig} 
\usepackage{hyperref}
\usepackage{graphics}
\usepackage{color}      
\usepackage{graphicx}
\usepackage{graphicx}
\usepackage{dcolumn}

\usepackage{amsmath} 
\begin{document}

\title{ A Universal Thermodynamic Inequality: Scaling Relations Between Current, Activity, and Entropy Production}
\author{Mesfin Asfaw  Taye}
\affiliation {West Los Angeles College, Science Division \\9000  Overland Ave, Culver City, CA 90230, USA}

\email{tayem@wlac.edu}

\begin{abstract}
We derive a universal thermodynamic bound constraining directional transport in both discrete and continuous nonequilibrium systems. For continuous-time Markov jump processes and overdamped diffusions governed by Fokker--Planck equations, we prove the inequality
$
\frac{2 V(t)^2}{A(t)} \leq \dot{e}_p(t),
$
linking the squared net velocity \( V(t) \), entropy production rate \( \dot{e}_p(t) \), and dynamical activity \( A(t) \). This relation captures a fundamental trade-off between transport, dissipation, and fluctuation intensity, valid far from equilibrium and without detailed balance. In addition, we introduce dimensionless thermodynamic ratios that quantify dissipation asymmetry, entropy extraction, and relaxation. These scaling laws unify discrete and continuous stochastic thermodynamics and provide experimentally accessible constraints on transport efficiency in nanoscale machines and active systems.
\end{abstract}

\pacs{Valid PACS appear here}
\maketitle


{\it Introduction.\textemdash}
Thermodynamic uncertainty relations (TURs) have become fundamental results in nonequilibrium statistical physics, revealing a universal trade-off between the precision of physical currents and the thermodynamic cost required to sustain them~\cite{mg1,mg2,mg3}. These inequalities impose lower bounds on entropy production based on the fluctuations of measurable quantities, such as particle flow, heat, or work. By relating  irreversibility to observable noise, TURs provide general constraints on the performance of small-scale systems that operate  far from equilibrium~\cite{mg1,mg2}. Initially formulated for steady-state stochastic processes~\cite{mg1}, TURs have been generalized to a wide variety of settings that include time-dependent driving, periodically modulated systems, and complex energy landscapes~\cite{mg4,mg5,mg6,mg7,mg8}.  As a result, they now serve as important  tools for probing nonequilibrium behavior across physical, chemical, and biological systems.

Brownian motors  as well as microscopic devices that harness thermal fluctuations to perform directed motion  often offer a particularly rich setting in which to explore the implications of TURs. These systems operate by breaking the detailed balance through spatial asymmetry or by  coupling to nonequilibrium reservoirs~\cite{mg9,mg10}. Early ratchet models showed that particles in asymmetric potentials under thermal gradients can produce net transport~\cite{mg12,mg13}. These pioneering works opened a path for extensive studies of systems with spatial variations in temperature, friction, and transition dynamics~\cite{mg14,mg15,mg16,c10,c11}  which in  turn  advances  our understanding of how nonequilibrium driving and structural asymmetry enable efficient transport at the microscale.

Despite extensive modeling efforts, a complete  understanding of how transport properties, such as velocity, dissipation, and efficiency, scale with system parameters remains incomplete.  Key questions persist: How do energy barriers, thermal asymmetry, and transition rate imbalances shape the trade-off between transport and thermodynamic costs?  Can general performance bounds be established independently of specific model details?

In this work, we establish a universal thermodynamic bound on directional transport that applies to both discrete-state and continuous nonequilibrium systems. The inequality
\[
\frac{2 V(t)^2}{A(t)} \leq \dot{e}_p(t)
\]
relates the squared current \( V(t) \), dynamical activity \( A(t) \), and entropy production rate \( \dot{e}_p(t) \) and holds far from equilibrium without assuming a detailed balance.  Unlike traditional thermodynamic uncertainty relations, this bound constrains instantaneous transport rather than fluctuations.  We verify its sharpness using a solvable three-state Brownian motor, revealing distinct transport regimes and efficiency scaling with barrier height, temperature asymmetry, and transition bias.  Our results provide a model-independent framework for quantifying transport--dissipation trade-offs in mesoscopic systems.

{\it Thermodynamic uncertainty–like scaling in discrete Brownian ratchets .\textemdash} Consider a continuous-time Markov process on a finite state space \( \{i\} \), with transition rates \( P_{ij}(t) \) and time-dependent state probabilities \( p_i(t) \). Define the instantaneous flux and current as
\begin{equation}
q_{ij}(t) := p_i(t) P_{ij}(t), \qquad
J_{ij}(t) := q_{ij}(t) - q_{ji}(t).
\end{equation}
We adopt antisymmetric summation \( \sum_{i > j} \) throughout the paper.

The reduced activity, summing jump fluxes in one direction per pair, is defined as
\begin{equation}
A'(t) := \sum_{i > j} p_i(t) P_{ij}(t).
\end{equation}
The total dynamical activity, summing all transitions across both directions, is then
$
A(t) := \sum_{i \ne j} p_i(t) P_{ij}(t) = 2 A'(t).
$

For a Brownian particle moving along a one- or two-dimensional discrete ratchet potential, the Boltzmann-Gibbs form of entropy,
\begin{equation}
S[p_i(t)] = -\sum_{i} p_i(t) \ln p_i(t),
\end{equation}
remains valid even in nonequilibrium conditions. The entropy extraction rate \( \dot{h}_d(t) \) can be expressed microscopically in terms of local probabilities and transition rates as
\begin{align}
\dot{h}_d(t) &= \sum_{i>j} \left[ p_i P_{ji} - p_j P_{ij} \right] \ln \left( \frac{P_{ji}}{P_{ij}} \right) \nonumber \\
&= \dot{e}_p(t) - \dot{S}(t),
\end{align}
where the total entropy production rate is
\begin{equation}
\dot{e}_p(t) = \sum_{i>j} \left[ p_i P_{ji} - p_j P_{ij} \right] \ln \left( \frac{p_i P_{ji}}{p_j P_{ij}} \right),
\end{equation}
and the system entropy change rate is
\begin{equation}
\dot{S}(t) = \sum_{i>j} \left[ p_i P_{ji} - p_j P_{ij} \right] \ln \left( \frac{p_i}{p_j} \right).
\end{equation}
Here, \( \dot{e}_p(t) \) and \( \dot{S}(t) \) quantify the irreversible entropy production and the time rate of change of system entropy, respectively.

The total net current across a chosen oriented cycle \( C \) is
\begin{equation}
V(t) := \sum_{(i \to j) \in C} J_{ij}(t).
\end{equation}
Now to   derive a thermodynamic uncertainty–like scaling, let us 
 write
$
x_{ij}(t) = \frac{J_{ij}(t)}{\sqrt{q_{ij}(t)}}$  and 
$y_{ij}(t) = \sqrt{q_{ij}(t)}$
so that \( J_{ij}(t) = x_{ij}(t) y_{ij}(t) \). The cycle current becomes
$
V(t) = \sum_{(i \to j) \in C} x_{ij}(t) y_{ij}(t).
$
Using the Cauchy--Schwarz inequality:
$
V(t)^2 \leq \left( \sum_{(i \to j) \in C} \frac{J_{ij}^2(t)}{q_{ij}(t)} \right)
\left( \sum_{(i \to j) \in C} q_{ij}(t) \right).
$
Extending to the full antisymmetric case
\begin{equation}
V(t)^2 \leq \left( \sum_{i > j} \frac{J_{ij}^2(t)}{q_{ij}(t)} \right) A'(t).
\end{equation}
We now apply the inequality
\begin{equation}
\frac{(a - b)^2}{a} \leq (a - b) \ln \left( \frac{a}{b} \right), \quad \forall\, a, b > 0,
\end{equation}
with \( a = q_{ij}(t),\, b = q_{ji}(t) \), one gets
\begin{equation}
\frac{J_{ij}^2(t)}{q_{ij}(t)} \leq J_{ij}(t) \ln \left( \frac{q_{ij}(t)}{q_{ji}(t)} \right).
\end{equation}
Summing over \( i > j \), we find
\begin{equation}
\sum_{i > j} \frac{J_{ij}^2(t)}{q_{ij}(t)} \leq \dot{e}_p(t).
\end{equation}
Substituting into the earlier equation
$
V(t)^2 \leq A'(t) \cdot \dot{e}_p(t)
$,
we write  the dimensionless inequality as 
\begin{equation}
\zeta(t) := \frac{2V(t)^2}{A(t) \cdot \dot{e}_p(t)} \leq  1
\end{equation}
in terms of the total activity \( A(t) = 2 A'(t) \). The dimensionless quantity $\zeta$ characterizes how efficiently directed motion is sustained relative to the thermodynamic cost and dynamical fluctuations. When $\zeta = 1$, the current is maximally organized and tightly coupled to dissipation, while $\zeta < 1$ reflects the presence of excess activity or noise not contributing to net transport.

The steady-state entropy production rate in a unicyclic Markov process with $N$ states is given by $\dot{e}_p = \frac{v}{N} \mathcal{F}$, where $v$ is the net velocity and $\mathcal{F} = \sum_{i=1}^N \ln(P_{i,i+1}/P_{i+1,i})$ is the total cycle affinity. A general bound constrains the current via $\zeta = 2v^2 / (A \dot{e}_p) \leq 1$, which is saturated when $v = \frac{1}{2N} A \mathcal{F}$. Equality holds trivially at equilibrium ($\mathcal{F} = 0$) but also in certain nonequilibrium regimes where entropy production is nonzero as shown in Fig. 1b. We verified this condition in a three-state model, finding \cite{c12} that the bound is exactly saturated ($\zeta = 1$) for specific parameter choices. Physically, $\zeta = 1$ indicates that dissipation is maximally and coherently converted into directed current, without excess dynamical activity or inefficiency.

The instantaneous inequality
$
\frac{2 V(t)^2}{A(t)} \leq \dot{e}_p(t)
$
can be extended to a finite time window \( [0, \tau] \) by defining time-averaged quantities:
\begin{equation}
\bar{V} := \frac{1}{\tau} \int_0^\tau V(t)\, dt, \quad 
\bar{A} := \frac{1}{\tau} \int_0^\tau A(t)\, dt, \quad 
\bar{e}_p := \frac{1}{\tau} \int_0^\tau \dot{e}_p(t)\, dt.
\end{equation}
From these, a finite-time bound follows:
\begin{equation}
\frac{2 \bar{V}^2}{\bar{A}} \leq \bar{e}_p,
\end{equation}
which relates cumulative current, activity, and entropy production. This result is experimentally accessible and applicable even in strongly driven regimes. Alternatively, integrating the pointwise inequality yields
$
\int_0^\tau \frac{2 V(t)^2}{A(t)}\, dt \leq \int_0^\tau \dot{e}_p(t)\, dt,
$
allowing for refined analysis of systems with time-varying dissipation or transport. 

The transport--dissipation inequality expresses a fundamental thermodynamic constraint: sustained directed motion incurs a cost in either entropy production or dynamical activity. Achieving a large current $V(t) $ necessitates increased irreversibility via $ {\dot e}_p(t)$ or more frequent transitions via $ A(t)$. As $ \zeta(t) \to 1 $, the dissipation and fluctuations are maximally converted into coherent motion; lower values indicate inefficiency in utilizing these resources. 

The transport--dissipation inequality provides a universal constraint on nonequilibrium dynamics, bounding directional current by entropy production, and dynamical activity. Generalizing thermodynamic uncertainty relations, it applies far from equilibrium and without a detailed balance. This relation enables entropy production to be inferred from measurable trajectory observables, offering a practical tool for quantifying irreversibility and efficiency in mesoscopic systems, such as molecular motors, colloidal ratchets, and nanoscale circuits.

We consider the exactly solvable three-state Brownian motor model introduced in Ref.~\cite{c12}, which allows exact analytical expressions for steady-state probabilities, current, entropy production, and dynamical activity. Using this model, we test the transport--dissipation--activity inequality and numerically evaluate the dimensionless parameter $\zeta$, confirming that the bound $\zeta \leq 1$ holds across all regimes. As shown in Fig.1, the inequality saturates to unity  near equilibrium, where current and entropy production vanish while activity remains finite, which is consistent with reversible dynamics.
\begin{figure}[ht]
\centering
\includegraphics[width=0.4\textwidth]{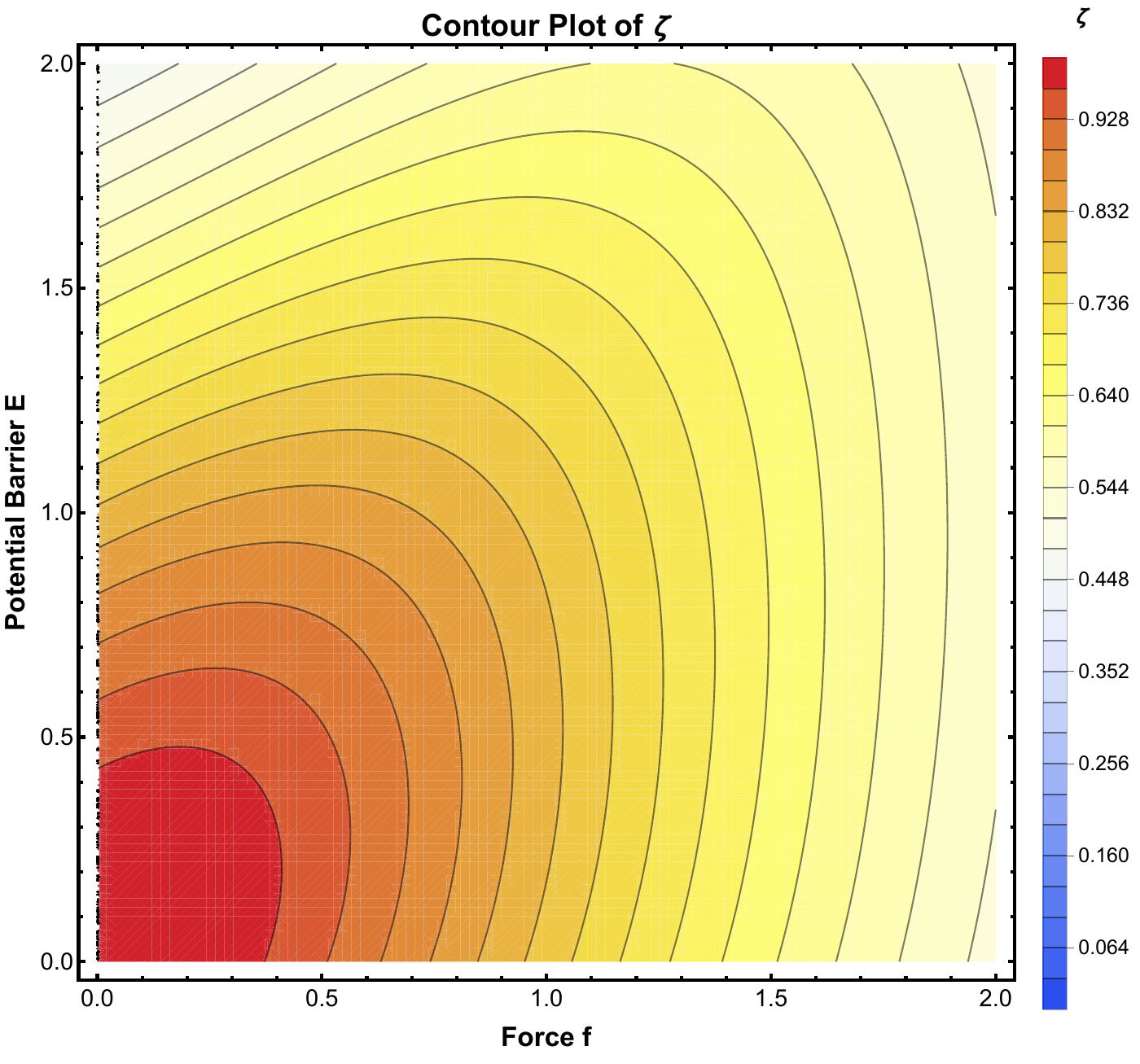}\hfill
\includegraphics[width=0.4\textwidth]{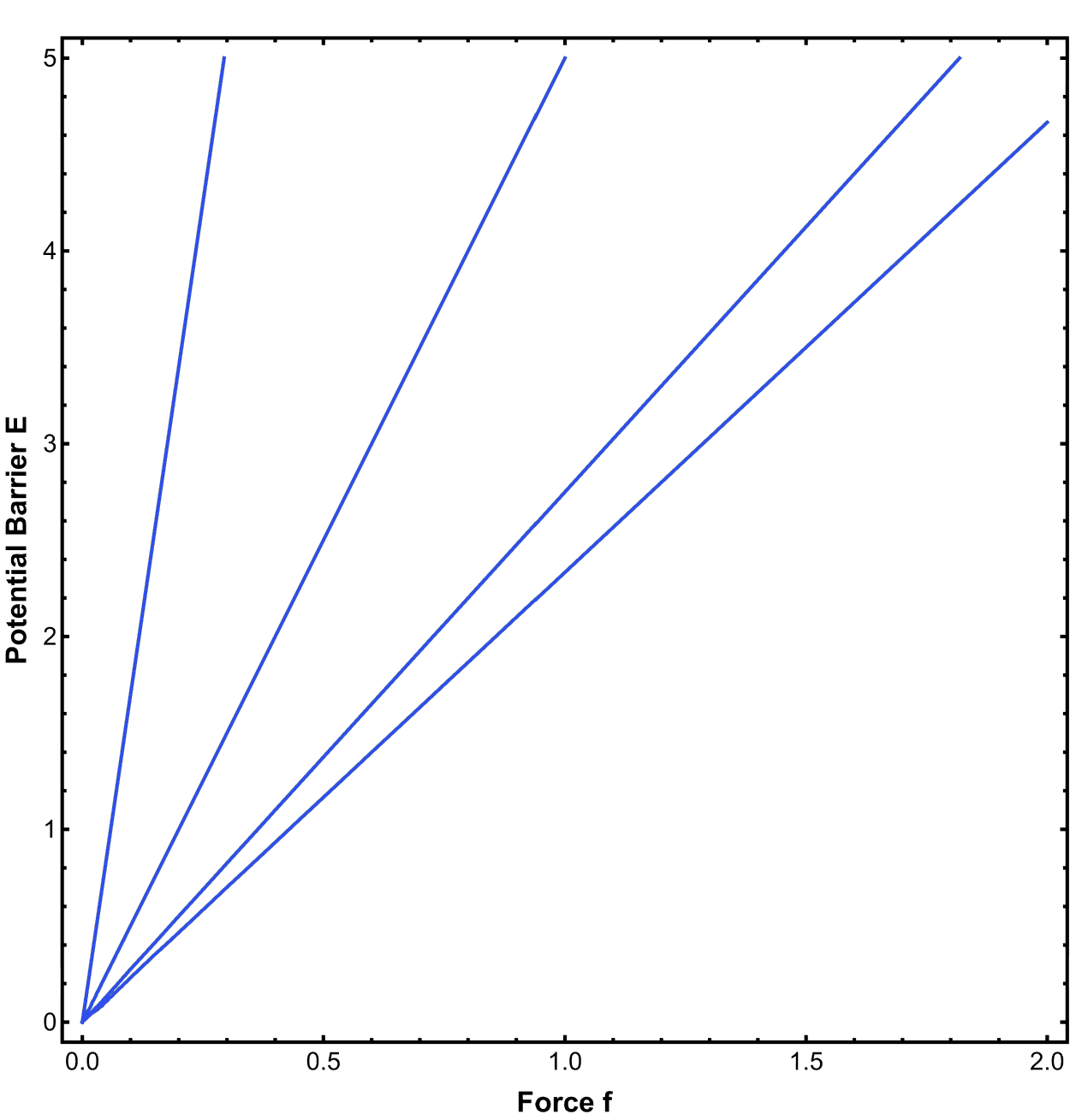}
\caption{
(a) Contour plot of steady-state transport efficiency $\zeta$ as a function of applied force $f$ and potential barrier height $E$ under isothermal conditions. The plot shows how $\zeta$ varies non-monotonically with both parameters, highlighting the interplay between driving strength and energy landscape in determining transport efficiency.
(b) Contour plot of the bound-saturating expression $v = \frac{1}{2N} A \mathcal{F}$ as a function of $E$ and $f$, with fixed temperature $T_h = 1.2, 2, 5, 10$ from left to right. This illustrates how the equality condition $\zeta = 1$ can emerge even out of equilibrium depending on the driving and thermal parameters.
}
\label{fig:zeta_combined}
\end{figure}

{\it Other Scaling Relations for a Brownian Heat Engine Between Hot and Cold Baths.\textemdash}  Let us now  introduce  scaling relation  in terms of  
 entropy extraction rate \( \dot{h}_d(t) \), the entropy production rate \( \dot{e}_p(t) \)   and   the rate of change of the system entropy \( \dot{S}(t) \). 

We define the dissipation-to-extraction ratio
\begin{equation}
\gamma(t) := \frac{\dot{e}_p(t)}{\dot{h}_d(t)},
\end{equation}
which measures the degree of internal irreversibility relative to environmental entropy extraction. In the steady state, where \( \dot{S}(t) = 0 \), this reduces to \( \gamma(t) = 1 \). Values \( \gamma(t) > 1 \) indicate entropy accumulation within the system, while \( \gamma(t) < 1 \) corresponds to internal ordering.

To connect thermodynamic cost to directed motion, we define the transport-normalized entropy extraction:
\begin{equation}
\theta(t) := \frac{V(t)^2}{\dot{h}_d(t)},
\end{equation}
which quantifies how much coherent transport is achieved per unit of entropy removed by the bath.

From the  universal bound
$
\frac{2 V(t)^2}{A(t)} \leq \dot{e}_p(t),
$
we obtain the scaling inequality
\begin{equation}
\theta(t) \leq \frac{1}{2} A(t)\, \gamma(t),
\end{equation}
indicating that high-performance transport requires either elevated entropy production or high fluctuation activity.

To quantify the system's deviation from steady state, we define the normalized entropy imbalance as 
\begin{equation}
\chi(t) := \frac{\dot{e}_p(t) - \dot{h}_d(t)}{\dot{e}_p(t) + \dot{h}_d(t)} = \frac{\dot{S}(t)}{\dot{e}_p(t) + \dot{h}_d(t)}.
\end{equation}
This dimensionless parameter  ratio vanishes at steady state and captures the degree of net entropy accumulation or release during transient evolution.

While the naive inequality
$
\dot{e}_p(t)^2 \geq \dot{S}(t)^2 + \dot{h}_d(t)^2
$
is generally invalid, we propose a sharpened bound incorporating transport,
\begin{equation}
\dot{e}_p(t)^2 \geq \dot{h}_d(t)^2 + \frac{2 V(t)^2}{A(t)},
\end{equation}
which reflects the irreducible thermodynamic cost required to sustain both entropy flow and coherent motion.

{\it Scaling Relations for Efficiency and COP of Refrigerators.\textemdash}
We next consider refined scaling relations for irreversible energy conversion in steady-state heat engines and refrigerators operating between two thermal reservoirs at temperatures \( T_h \) and \( T_c \). Let \( \dot{Q}_h \) and \( \dot{Q}_c \) denote heat flows from the hot and to the cold reservoirs, respectively, with \( f \) an external force driving motion at mean velocity \( V \), and \( \dot{e}_p \) the steady-state entropy production rate.

The entropy balance relation reads
\begin{equation}
\dot{e}_p = \frac{\dot{Q}_c}{T_c} - \frac{\dot{Q}_h}{T_h},
\end{equation}
while the first law yields the useful power output
$
\dot{W} = f V = \dot{Q}_h - \dot{Q}_c
$.
Eliminating \( \dot{Q}_c \) leads to a power-entropy relation:
$
\dot{e}_p = \dot{Q}_h \left( \frac{1}{T_c} - \frac{1}{T_h} \right) - \frac{f V}{T_c},
$
which rearranges as
$
f V + T_c \dot{e}_p = \dot{Q}_h \left( 1 - \frac{T_c}{T_h} \right).
$
This expression reveals how total dissipation constrains extractable power under finite heat absorption.

The efficiency of the engine is defined by
$
\eta := \frac{f V}{\dot{Q}_h},
$
so that
\begin{equation}
\eta = \left( 1 - \frac{T_c}{T_h} \right) - \frac{T_c}{\dot{Q}_h} \dot{e}_p.
\end{equation}
Thus, the deviation from Carnot efficiency is proportional to entropy production per unit absorbed heat. A normalized efficiency ratio is
\begin{equation}
\Gamma := \frac{\eta}{\eta_C} = 1 - \frac{T_c \dot{e}_p}{\dot{Q}_h \left( 1 - \frac{T_c}{T_h} \right)},
\end{equation}
which satisfies
$
0 \leq \Gamma < 1
$,
with equality only in the quasistatic limit \( \dot{e}_p \to 0 \).

Analogous reasoning applies to nonequilibrium refrigerators. Let \( Q_c \) denote the heat extracted from the cold reservoir, and \( f V \) the work input. The coefficient of performance (COP) is defined as
$
\text{COP} := \frac{Q_c}{f V}.
$
The Carnot limit reads
$
\text{COP}_{\text{Carnot}} = \frac{T_c}{T_h - T_c},
$
and thermodynamic consistency requires
$
\text{COP} < \text{COP}_{\text{Carnot}}
$.
After  some algebra, we get 
\begin{equation}
\dot{e}_p = Q_c \left( \frac{1}{T_c} - \frac{1}{T_h} \right) - \frac{f V}{T_h}.
\end{equation}
One can also solve  \( Q_c \)  as 
$
Q_c = \left( \frac{T_c T_h}{T_h - T_c} \right) \left( \dot{e}_p + \frac{f V}{T_h} \right),
$
which inserted into the COP yields
\begin{equation}
\text{COP} = \left( \frac{T_c T_h}{T_h - T_c} \right) \left( \frac{\dot{e}_p}{f V} + \frac{1}{T_h} \right).
\end{equation}
Finally, defining the normalized ratio
\begin{equation}
\Gamma_{\text{COP}} := \frac{\text{COP}}{\text{COP}_{\text{Carnot}}}
= 1 - \frac{T_h \dot{e}_p}{Q_c},
\end{equation}
we again obtain \( \Gamma_{\text{COP}} < 1 \), with equality only when \( \dot{e}_p \to 0 \).

These relations quantify how entropy production fundamentally limits efficiency and performance in both energy extraction and refrigeration, and connect entropy budgets to observable currents and thermodynamic cost.

{\it Thermodynamic Uncertainty–Like Scaling in Continuum Brownian Ratchets.\textemdash}
  We now derive the continuum analog of the current–activity–dissipation inequality for overdamped diffusion in \( d \)-dimensional space. The dynamics are governed by the Fokker--Planck equation
\begin{equation}
\frac{\partial P(x,t)}{\partial t} = -\nabla \cdot \mathbf{J}(x,t),
\end{equation}
with the probability current
$
\mathbf{J}(x,t) = \mathbf{F}(x)\, P(x,t) - D(x)\, \nabla P(x,t),
$
where \( \mathbf{F}(x) \) is an external drift and \( D(x) \) is a (possibly position-dependent) diffusivity. We assume unit mobility without loss of generality.

To derive the bound, we define the entropy production rate
\begin{equation}
\dot{e}_p(t) = \int dx\, \frac{|\mathbf{J}(x,t)|^2}{D(x) P(x,t)},
\end{equation}
the symmetric dynamical activity
$
A'(t) = \frac{1}{2} \int dx\, D(x) P(x,t),
$
and the net transport
$
V(t) = \left| \int dx\, \mathbf{J}(x,t) \right|.
$
Introduce the auxiliary functions
\begin{equation}
f(x) = \frac{\mathbf{J}(x,t)}{\sqrt{D(x) P(x,t)}}, \quad 
g(x) = \sqrt{\tfrac{1}{2} D(x) P(x,t)},
\end{equation}
so that
\begin{equation}
V(t) = \left| \int dx\, \mathbf{J}(x,t) \right| 
= \sqrt{2} \left| \int dx\, f(x) g(x) \right|.
\end{equation}
Applying the Cauchy--Schwarz inequality yields
$
V(t)^2 \leq 2 \left( \int dx\, f(x)^2 \right) \left( \int dx\, g(x)^2 \right),
$
which simplifies to
$
V(t)^2 \leq 2\, \dot{e}_p(t) \cdot A'(t).
$
We thus obtain the final scaling inequality
\begin{equation}
\boxed{ \frac{2 V(t)^2}{A(t)} \leq \dot{e}_p(t) }.
\end{equation}
We retrieved the discrete-state result in the continuum limit, confirming the universality of the transport–dissipation–activity relation.

{\it Summary and conclusion.\textemdash}
We have established a general thermodynamic scaling framework that constrains transport in nonequilibrium discrete-state and continuum Brownian systems. A central result is a universal inequality linking net current, entropy production, and dynamical activity—extending thermodynamic uncertainty principles beyond near-equilibrium and detailed balance regimes. Applied to a solvable three-state Brownian motor, the bound is analytically and numerically validated, with saturation occurring only at reversibility. Additional scaling relations connect entropy extraction, system entropy change, and energetic efficiency, revealing intrinsic trade-offs between coherence, dissipation, and fluctuation. These results provide a compact, model-independent foundation for quantifying performance limits in mesoscopic thermal machines, and enable dissipation to be inferred from measurable trajectory statistics, offering broad utility in theory, simulation, and experiment.

\section*{Acknowledgment}
I would like to thank  Mulu  Zebene and Asfaw Taye for the
constant encouragement. 

\section*{Data Availability Statement }This manuscript has no
associated data or the data will not be deposited. [Authors’
comment: Since we presented an analytical work, we did not
collect any data from simulations or experimental observations.]

\end{document}